\documentclass[10pt,onecolumn,amsmath,amssymb]{article}
\usepackage{amsfonts}
\usepackage{amsmath}
\usepackage{amssymb}
\usepackage{graphicx}
\usepackage{bm}
\usepackage[super]{cite}
\usepackage{times}

\usepackage[font={small}]{caption}

\newcommand{\onlinecite}[1]{\citen{#1}}
\newcommand{\refcite}[1]{\onlinecite{#1}}

\makeatletter
\renewcommand\section{\@startsection{section}{1}{\z@}%
{-3.5ex \@plus -1ex \@minus -.2ex}%
{2.3ex \@plus.2ex}%
{\normalfont\bfseries}}
\renewcommand\subsection{\@startsection{subsection}{1}{\z@}%
{-3.5ex \@plus -1ex \@minus -.2ex}%
{2.3ex \@plus.2ex}%
{\normalfont\bfseries}}
\makeatother

\date{}

\renewcommand{\v}[1]{|{#1}\rangle}
\newcommand{\iv}[1]{\langle{#1}|}
\newcommand{\abs}[1]{\left| {#1}\right|}

\newcommand{\conj}{^{*}}
\newcommand{\expi}[1]{e^{i\left({#1}\right)}}

\begin{document}

\title{\large\bf
Reduction of branching graphs
\\\large\bf
supporting continuous time return quantum walks}

\author{\normalsize
Thomas Cavin
\\\normalsize\it
Department of Physics, St. Louis University, St. Louis, Missouri  63103, USA
\\
\\\normalsize
Dmitry Solenov
\\\normalsize\it
Department of Physics, St. Louis University, St. Louis, Missouri  63103, USA
\\\normalsize\it
solenov@slu.edu
}

\maketitle\thispagestyle{empty}

\begin{abstract}
We demonstrate that continuous time quantum walks on several types of branching graphs, including graphs with loops, are identical to quantum walks on simpler linear chain graphs. We also show graph types for which such equivalence does not exist. Several instructive examples are discussed, and a general approach to analyze more complex branching graphs is formulated. It is further illustrated with a return quantum walk solution for a cube graph with adjustable complex hopping amplitudes.
\end{abstract}

\section{Introduction}

A continuous time quantum walk\cite{Farhi,Childs,Kempe,Tregenna,Kendon,Solenov-W1,Solenov-W2,Solenov-W3,Mulken} is quantum propagation through a discrete set of states, a graph, in time. Edges of the graph correspond to amplitudes for a quantum particle to transition between the connected states. All amplitudes in the graph define the adjacency matrix, or a Hamiltonian, that governs the evolution in that quantum system. The quantities of interest for quantum walks are some characteristic states of the system and times when they are achieved, including hitting time,\cite{Aharanov} mixing time,\cite{Aharanov,Solenov-W1,Solenov-W2,Solenov-W3} or return time.\cite{solenov-3qb,solenov-gates} The return time, for example, is of particular interest for constructing quantum gates\cite{solenov-3qb,solenov-gates,DiVincenzo,Chakrabarti,nielsenchuang} in system with auxiliary states.\cite{solenov-3qb,Solenov-2QB,Solenov-NV,Solenov-QD,Solenov-QDs}

In principle, the problem of determining quantum evolution through a finite coherent system of distinct quantum states is equivalent to the problem of diagonalization of the Hamiltonian of such system, which can be done numerically if the number of states is not prohibitively large. Some smaller or highly symmetric systems can be solved analytically.\cite{Solenov-W1,Solenov-W2,Solenov-W3,solenov-gates,Moore,Alagic} At the same time, optimization or identification of characteristic quantities of interest to continuous time quantum walks can be numerically challenging even for relatively small systems, which can be due, e.g., complexity of pattern recognition in these systems. In particular, the problem that arises, e.g., when constructing quantum gates based on quantum walks,\cite{solenov-gates} involves a search for times of complete, but non-trivial, return of population to the initial node in parametric space of all amplitudes in the adjacency matrix with some constraints. For this reason it is desirable to be able to analytically understand ``topological'' equivalence between graphs of different structure that are capable of performing the same type of quantum walks. It is important to stress that in the context of this investigation topological equivalence will refer to the ability to perform the same type of quantum walk and {\it not} to geometrical topological equivalence or topology of connections.

In this paper we investigate typical non-trivial graph elements (sub-graphs) that are expected\cite{solenov-3qb,solenov-gates} in three-qubit or larger systems with at least one actively used auxiliary state per qubit. We demonstrate topological equivalence between these elements and linear chain graphs, which are often approachable analytically.\cite{solenov-gates} We investigate branching graphs with short one-segment branches and longer two-segment branches, as well as graphs with single or multiple loops. We show that topological equivalence in a sense of quantum walks differs substantially from topological similarity in connectivity. For example, we demonstrate that a linear chain graph with two-segment branch (no loops) supports the same type of quantum walk as the linear chain graph with one four-segment loop. The results derived in this paper will help to formulate more efficient types of quantum gates for quantum computing for different architectures. Continuous time quantum walk on graphs with complex adjacency matrices\cite{Zimboras} are also of interest in relation with effects involving time-reversal symmetry breaking dynamics\cite{Peierls,Hofstadter,Sarma,Hasan,Dalibard}, exciton transport,\cite{Harel} and pattern recognition.\cite{Ambainis,Emms} The results presented in this paper can contribute to understanding of quantum transport and related phenomenta in these areas.

The paper is organized as follows: In Sec.~\ref{sec:1s} we investigate the simplest type of a non-trivial modification of linear chain graphs---addition of a single one-segment branch. We demonstrate that the system with one such branch remains topologically equivalent to a (one segment longer) linear chain graph in quantum sense as described above. The two systems are isomorphic in this sense. Specifically, any continuous time quantum walk that begins from a node below the attachment on the branched graph, or from equivalent node on a longer linear graph, produces exactly the same evolution of all amplitudes below the attachment as well as on half of the nodes above the attachment (including on the attachment vertex). This is demonstrated by introducing local iterative transformations of basis states to reduce the graph. These transformations can be performed analytically and do not affect quantum evolution through the system, hence providing simple analytical solutions to quantum walks in systems with one-segment branches.
Section~\ref{sec:3l} is devoted to discussion of a linear chain graph that contains one edge-sharing three-segment loop. This graph can be reduced to a linear chain graph at the cost of introducing self-loops to the system (diagonal entries in the adjacency matrix), which occurs because the original graph is not a bipartite graph. The non-bipartite nature of the original graph is thus preserved by the transformation. The graphs before and after the transformation remain topologically equivalent in quantum evolution sense.
In Sec.~\ref{sec:4l} we derive a transformation and demonstrate topological similarity (in quantum sense) between a linear chain graph with an edge-sharing four-segment loop and a longer linear chain graph. The two, however, are not always isomorphic, that is, a linear chain of certain length is equivalent to a chain with edge-sharing four-segment loop, but the latter is not always equivalent to the former. The condition on hopping amplitudes when the systems become isomorphic is derived. In Sec.~\ref{sec:romb}, we perform similar investigation of a linear chain graph with a four-segment loop incorporated diagonal-wise. We show that such graphs are topologically equivalent to graphs with a single one-segment branch, and, hence, to one-segment-longer linear chain graphs. In Sec.~\ref{sec:2s} we study linear graphs interrupted with a larger six-node loop (inserted symmetrically) and linear chain graphs with a single two-segment branch. We demonstrate that such graphs are equivalent to graphs with a four-segment loop investigated in Sec.~\ref{sec:4l}, and are conditionally equivalent to longer linear chain graphs. In Sec.~\ref{sec:Simp} we outline a general reduction approach  for more complex branching graphs. In Sec.~\ref{sec:ExCube} we apply this approach and investigate a multi-loop cube graph, which is of particular interest in constructing multiqubit quantum gates via auxiliary states.\cite{solenov-gates} Specifically, we demonstrate that, under certain conditions, a non-trivial cube graph can be transformed into two unconnected linear chain graphs with four nodes each. This makes analytical solution to quantum walks accessible in such systems. An example for some choice of amplitudes is given.

\section{One-segment branch}\label{sec:1s}

In this section we introduce a simple analytical procedure to transform a linear graph with a single-segment branch and arbitrary complex hopping amplitudes (edges) into a linear graph without branching. We first investigate the situation when only one branch is present and focus on graph nodes in the vicinity of that branch. We introduce a basis rotation that shifts the one-segment branch two nodes along the chain and show how all three involved connections (hopping amplitudes) are adjusted by this change. This procedure can be applied iteratively to move the one-segment branch to the end of the linear chain, hence analytically transforming the system into a linear chain graph\cite{solenov-gates} without branching. In the case when, after iterative application of the above procedure, the branch is still one node away from the end of the linear chain graph, rotation described in Sec.~V of Ref.~\refcite{solenov-gates} can be used to complete the transformation to a linear chain graph without branches. The total number of available free parameters (amplitudes) remains unchanged. When several non-connected branches are present, the procedure can be applied iteratively. The cases of connected branches (loops) are addressed in the next sections.

\begin{figure}\begin{center}
\includegraphics[width=0.3\textwidth]{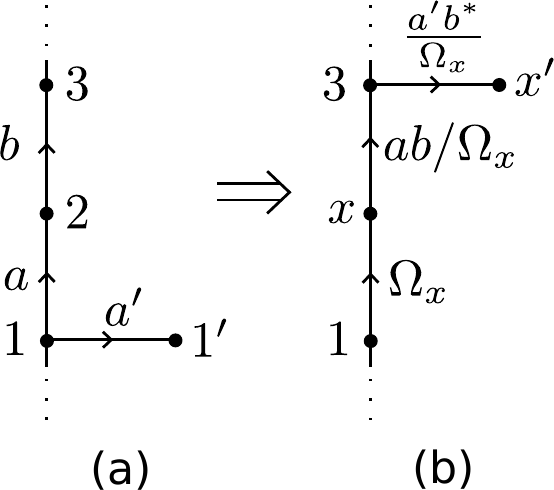}
\caption{\label{fig:one-branch}
A basis rotation which shifts a single-segment branch along the linear chain graph. Here $\Omega_x=\sqrt{|a|^2+|a'|^2}$. Points are graph vertices (states), each connection represents two hermitian-conjugate terms in the adjacency matrix connecting two states. Arrow indicate which of the two conjugated terms is displayed next to the line, e.g., in part (a), $1\leftrightarrow 2$ segment corresponds to terms $a\v{1}\iv{2}+a^{*}\v{2}\iv{1}$ (the other direction for the arrow would correspond to $a\v{2}\iv{1}+a^{*}\v{1}\iv{2}$).
}
\end{center}\end{figure}

The total Hamiltonian of the entire graph can be formulated as
\begin{eqnarray}\label{eq:1s:H}
H\{H_{ij}\} = ... + \xi\v{...}\iv{i} + H_{ij} + \xi' \v{j}\iv{...} + ...
\end{eqnarray}
where the part of the Hamiltonian describing graph in Fig.~\ref{fig:one-branch}(a)
\begin{eqnarray}\label{eq:1s:H13}
H_{13} = a \v{1}\iv{2} + a' \v{1}\iv{1'} + b \v{2}\iv{3} + h.c.
\end{eqnarray}
is connected to the remainder of the graph via nodes 1 and 3 (as indicated by the subscripts).

The first step in the transformation is to define two orthogonal states $\v{x}$ and $\v{x'}$ as
\begin{eqnarray}\label{eq:1s:xxp}
\left\{\begin{array}{l}
\v{x} = \frac{a^*\v{2} + a'^*\v{1'}}{\Omega_x}
\\
\v{x'} = \frac{a'\v{2} - a\v{1'}}{\Omega_x}
\end{array}\right.
\end{eqnarray}
with
\begin{eqnarray}\label{eq:1s:Omega_x}
\Omega_x = \sqrt{|a|^2+|a'|^2}.
\end{eqnarray}
This transformation corresponds to rotation of basis for subgraph containing states $\v{2}$ and $\v{1'}$, and replaces the first two terms in Eq.~(\ref{eq:1s:H13}) with $\Omega_x\v{1}\iv{x}$. The last term splits into two, because 
\begin{eqnarray}\label{eq:1s:2}
\v{2} = \frac{a\v{x} + a'^*\v{x'}}{\Omega_x}
\end{eqnarray}
The overall Hamiltonian of the subgraph $H_1$ becomes
\begin{eqnarray}\label{eq:1s:H1:new}
H_{13} = \Omega_x \v{1}\iv{x} 
+ \frac{ab}{\Omega_x} \v{x}\iv{3} 
+ \frac{a'b^*}{\Omega_x} \v{3}\iv{x'} 
+ h.c.
\end{eqnarray}
The corresponding graph is shown in Fig.~\ref{fig:one-branch}(b). Note that Hamiltonian (\ref{eq:1s:H}) remains unchanged except for $H_{13}$.

\section{Edge-sharing three-segment loop}\label{sec:3l}

\begin{figure}\begin{center}
\includegraphics[width=0.3\textwidth]{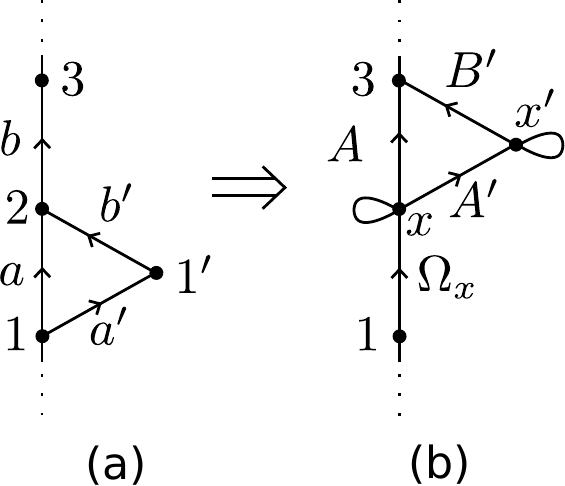}
\caption{\label{fig:3l}
The transformation of an edge-sharing three-segment loop subgraph. Graphic notations are the same as in Fig.~\ref{fig:one-branch}. Self-loops denote diagonal entires in the adjacency matrix (Hamiltonian).}
\end{center}\end{figure}

In this section we demonstrate how a graph that contains a three-segment edge-sharing loop [see Figure~\ref{fig:3l}(a)] can be iteratively transformed into a linear chain graph. This transformation does not affect quantum walks that begin on any node below (and including) node 1. Unlike in the previous case, however, the reduction procedure introduces diagonal terms to the adjacency matrix (self-loops in the graph). This occurs due to the fact that this graph [Figure~\ref{fig:3l}(a)] is not a bipartite graph, in contrast to other graphs discussed in this paper.

The Hamiltonian corresponding to graph shown in Figure~\ref{fig:3l}(a) is
\begin{eqnarray}\label{eq:3l:H}
H_{13} = 
a \v{1}\iv{2} 
+a' \v{1}\iv{1'} 
+b' \v{1'}\iv{2} 
+b \v{2}\iv{3} 
+ h.c.
\end{eqnarray}
The first and the second terms can be combined by introducing rotation (\ref{eq:1s:xxp}) as before, with $\Omega_x$ given by Eq.~(\ref{eq:1s:Omega_x}). This transformation rotates the third term of the Hamiltonian as
\begin{eqnarray}\label{eq:3l:1p2}
\v{1'}\iv{2} = 
\frac{a'a^*}{\Omega_x^2}(\v{x}\iv{x}-\v{x'}\iv{x'})
+\frac{|a'|^2-|a|^2}{\Omega_x^2}\v{x}\iv{x'}
\end{eqnarray}
and the fourth term as
\begin{eqnarray}\label{eq:3l:23}
\v{2}\iv{3} = 
\frac{a}{\Omega_x}\v{x}\iv{3}
+\frac{a'^*}{\Omega_x}\v{x'}\iv{3}
\end{eqnarray}
After the transformation, the Hamiltonian takes the form
\begin{eqnarray}\label{eq:3l:H-new}
H_{13} \!=\! 
\Omega_x \v{1}\iv{x} 
\!+\!A \v{x}\iv{3} 
\!+\!A'\v{x}\iv{x'} 
\!+\!B'\v{x'}\iv{3}
\!+\!{\cal E}_x\v{x}\iv{x}
\!+\!{\cal E}_{x'}\v{x'}\iv{x'}
\!+\! h.c.
\end{eqnarray}
with
\begin{eqnarray}\label{eq:3l:AAB}
\left\{\begin{array}{l}
A = ab/\Omega_x
\\
A'= b'\frac{|a'|^2-|a|^2}{\Omega_x^2}
\\
B'= a'^*b/\Omega_x
\\
{\cal E}_x = a^*a'b'/\Omega_x^2 = -{\cal E}_{x'}
\end{array}\right.
\end{eqnarray}
The corresponding graph is shown in Fig.~\ref{fig:3l}(b).
Note that if a self-loop is present on node $1$ in the original graph [term $\sim\v{1}\iv{1}$ in the Hamiltonian (\ref{eq:3l:H})], it remains unaffected by the transformation. On the other hand, diagonal term $E_{1'}\v{1'}\iv{1'}$, if present in Eq.~(\ref{eq:3l:H}), alters amplitudes as
\begin{eqnarray}\label{eq:3l:AtoA}
&&A'\to A' - E_{1'} aa'/\Omega_x
\\\label{eq:3l:EtoE}
&&{\cal E}_x\to {\cal E}_x + |a'|^2/2\Omega_x
\\\label{eq:3l:EptoEp}
&&{\cal E}_{x'}\to {\cal E}_{x'} + |a|^2/2\Omega_x
\end{eqnarray}
Other parts of larger Hamiltonian $H\{H_{13}\}$ remain unchanged. This procedure shifts the loop one segment up the chain and can eliminate it if applied iteratively. Note that diagonal entries are introduced for each state above and including the initial state $\v{1}$. These self-loops preserve the non-bipartite nature of the original graph.

\section{Edge-sharing four-segment loop}\label{sec:4l}

\begin{figure}\begin{center}
\includegraphics[width=0.99\textwidth]{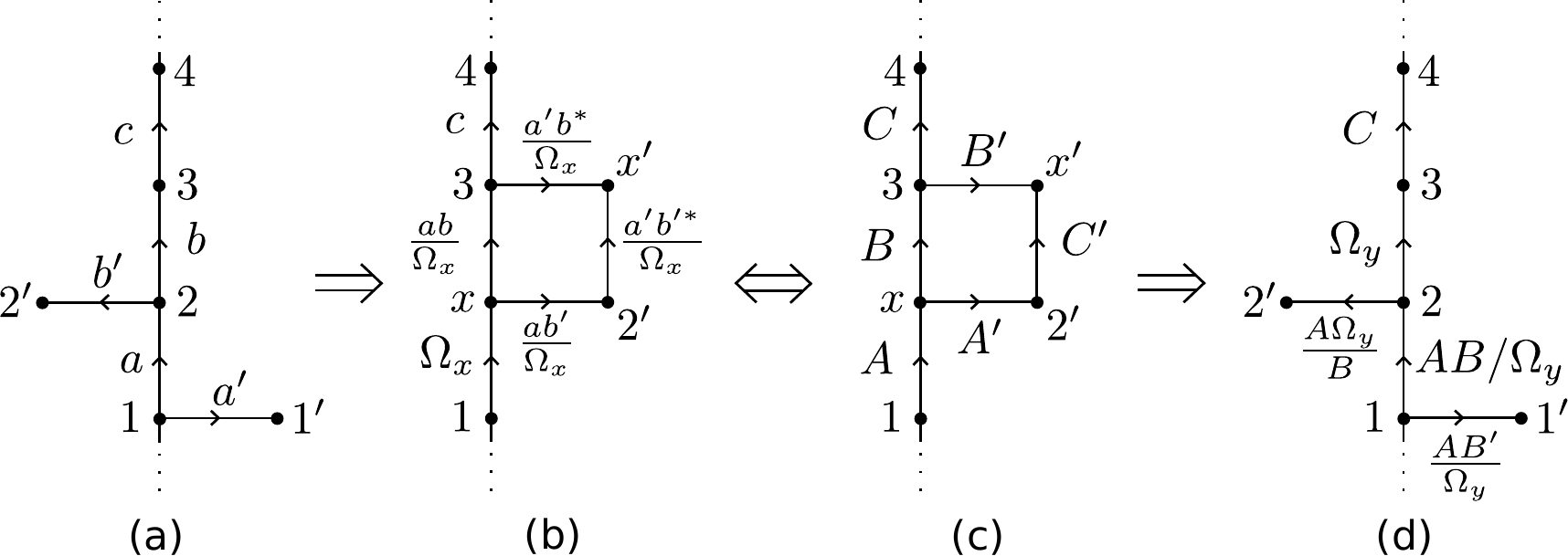}
\caption{\label{fig:4-loop}
The transformation between a four-segment loop subgraph and a subgraph with two one-segment branches. Here $\Omega_y = \sqrt{|B|^2+|B'|^2}$ and other notations are the same as in Fig.~\ref{fig:one-branch}. Note that transformation from (c) to (d) is possible only when condition (\ref{eq:4l:cond}) is satisfied.
}
\end{center}\end{figure}

In this section we derive a transformation to reduce a four-segment loop that shares one edge with the linear chain graph on which it is located. We start with two single-segment branches attached to two adjacent nodes of a linear chain graph [see Fig.~\ref{fig:4-loop}(a)]. The Hamiltonian corresponding to this piece of the graph is
\begin{eqnarray}\label{eq:4l:H14}
H_{14} = a\v{1}\iv{2} + a'\v{1}\iv{1'}
+ b\v{2}\iv{3} + b'\v{2}\iv{2'}
+c\v{3}\iv{4} + h.c.
\end{eqnarray}
and the adjacency matrix of the entire graph is given by Eq.~(\ref{eq:1s:H}). The transformation (\ref{eq:1s:xxp}) obtained in the previous section can be applied to states $\v{2}$ and $\v{1'}$ as before. Now, however, state $\v{2}$ is also connected to $\v{2'}$ and due to Eq.~(\ref{eq:1s:2}) we obtain
\begin{eqnarray}\label{eq:4l:H14:new}
H_{14}
\!=\!
\Omega_x\v{1}\iv{x} 
\!+\!
\frac{ab}{\Omega_x}\v{x}\iv{3}
\!+\!
\frac{b^*a'\!\!}{\Omega_x}\v{3}\iv{x'} 
\!+\!
\frac{ab'\!\!}{\Omega_x}\v{x}\iv{2'} 
\!+\!
\frac{b'a'^*\!\!\!}{\Omega_x}\v{x'}\iv{2'} 
\!+\!
c\v{3}\iv{4} 
\!+\!
h.c.
\end{eqnarray}
The corresponding graph, shown in Fig.~\ref{fig:4-loop}(b), is a four-segment loop attached to a linear chain.

The reversed transformation can also be performed if we start with a linear chain graph with a four-segment loop attached, as in Fig.~\ref{fig:4-loop}(c). This leads to a set of equations
\begin{eqnarray}\label{eq:4l:rel}
\left\{\begin{array}{rcl}
C &=& c
\\
A &=& \Omega_x
\\
A' &=& \frac{ab'}{\Omega_x}
\\
B &=& \frac{ab}{\Omega_x}
\\
B' &=& \frac{a'b^*}{\Omega_x}
\\
C' &=& \frac{a'b'^*}{\Omega_x}
\end{array}\right.
\end{eqnarray}
The loop graph [see Fig.~\ref{fig:4-loop}(c)] has one extra complex parameter as compared to graph with two branches in Figs.~\ref{fig:4-loop}(a) and (d). Therefore transformation from a loop graph to a graph with two branches [see Figs.~\ref{fig:4-loop}(c) and (d)] can only occur if the number of free parameters is reduced. The necessary condition  can be derived by comparing equations in system (\ref{eq:4l:rel}). This yields
\begin{eqnarray}\label{eq:4l:cond}
\frac{B'}{B^*}
=
\frac{C'}{A'^*}
\end{eqnarray}
The rest of the solution is
\begin{eqnarray}\label{eq:4l:abc}
\left\{\begin{array}{rcl}
a &=& \frac{AB}{\Omega_y}
\\
a' &=& \frac{AB'}{\Omega_y}
\\
b &=& \Omega_y = \sqrt{|B|^2+|B'|^2}
\\
b' &=& \frac{A'}{B}\Omega_y
\\
c&=&C
\end{array}\right.
\end{eqnarray}
Note that single-segment branches [see Fig.~\ref{fig:4-loop}(d)] can be moved to one of the ends of the linear chain, one after the other, by iteratively applying transformation derived in Sec.~\ref{sec:1s}. Note also that there is another, trivial, solution $C'=0$, which is not of interest here. 
As the result, we see that continuous time quantum walk on a linear graph with the attached non-trivial four-segment loop shown in Figs.~\ref{fig:4-loop}(b) or (c) is {\it topologically equivalent} to a walk on a longer linear chain graph if condition (\ref{eq:4l:cond}) or the one related to it by symmetry is satisfied. Note that the number of free parameters is reduced by one condition during the transformation. This derivation does not prove that there are no other conditions on the loop amplitudes that make that graph topologically indistinguishable from a linear chain graph (in terms of quantum walks). Nevertheless, it appears that other transformations of basis involving nodes of the loop do not simplify it to disconnected single-segment branches. Note also that continuous time quantum walks on graphs shown in Fig.~\ref{fig:4-loop} do not depend on phases of the hopping amplitudes (see Ref.~\refcite{solenov-gates} for examples). For this reason $\Omega_y$, which can in principle have arbitrary phase, was chosen to be a real number.

\section{Four-segment rhomboidal insertion}\label{sec:romb}

\begin{figure}\begin{center}
\includegraphics[width=0.4\textwidth]{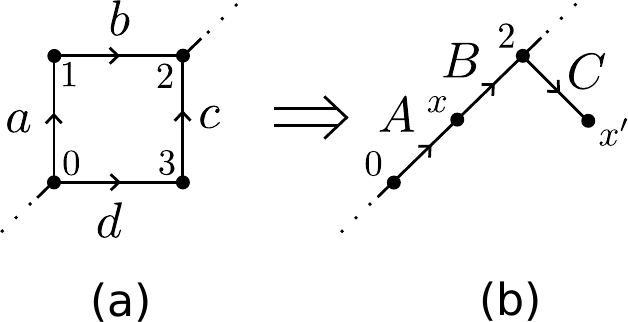}
\caption{\label{fig:romb}
The transformation between four-segment loop breaking a linear chain graph and a single-branch graph. Graphic notations are the same as in Fig.~\ref{fig:one-branch}.
}
\end{center}\end{figure}

In this section we derive transformation that changes a subgraph with a four-segment square attached via diagonally-opposite nodes into a linear chain graph with one single-segment branch as shown in Fig.~\ref{fig:romb}. The graphs remain topologically equivalent in term of quantum propagation, i.e., continuous time quantum walk is the same for both structures.

The Hamiltonian of the four-segment square part shown in Fig.~\ref{fig:romb}(a) is
\begin{eqnarray}\label{eq:romb:H02}
H_{02} = a \v{0}\iv{1} + b \v{1}\iv{2} 
+ c \v{3}\iv{2} + d \v{0}\iv{3} + h.c.
\end{eqnarray}
and the adjacency matrix of the entire graph has the form  of Eq.~(\ref{eq:1s:H}).
The first and the fourth terms can be combined to define a new state $\v{x}$ and state $\v{x'}$ orthogonal to one another,
\begin{eqnarray}\label{eq:romb:xxp}
\left\{\begin{array}{rcl}
\v{x}&=&\frac{a^*\v{1} + d^*\v{3}}{\Omega_x}
\\
\v{x'}&=&\frac{d\v{1} - a\v{3}}{\Omega_x}
\end{array}\right.
\\\label{eq:romb:Omega_x}
\Omega_x = \sqrt{|a|^2+|d|^2}
\end{eqnarray}
The remaining two terms of Eq.~(\ref{eq:romb:H02}), i.e. the second and the third terms, are transformed into this new basis as
\begin{eqnarray}\label{eq:romb:12}
b\v{1}\iv{2}&\to&
 \frac{ab}{\Omega_x}\v{x}\iv{2}
+ \frac{bd^*}{\Omega_x}\v{x'}\iv{2}
\\\label{eq:romb:32}
c\v{3}\iv{2}&\to&
 \frac{cd}{\Omega_x}\v{x}\iv{2}
- \frac{a^*c}{\Omega_x}\v{x'}\iv{2}
\end{eqnarray}
As the result, we obtain
\begin{eqnarray}\label{eq:romb:H02new}
H_{02} = A \v{0}\iv{x} + B \v{x}\iv{2} 
+ C \v{2}\iv{x'} + h.c.
\end{eqnarray}
with
\begin{eqnarray}\label{eq:romb:ABC}
\left\{\begin{array}{rcl}
A&=& \Omega_x
\\
B&=& \frac{ab+cd}{\Omega_x}
\\
C&=& \frac{b^*d-ac^*}{\Omega_x}
\end{array}\right.
\end{eqnarray}
which is the adjacency matrix for the graph shown in Fig.~\ref{fig:romb}(b). The one-segment branch on this graph can be pushed towards one of the ends to produce a linear chain graph as described in Sec.~\ref{sec:1s}. Note that the number of parameters (graph edges) is reduced from four to three complex numbers and further to three real numbers because quantum walk on a linear chain graph depends only on the absolute values of the amplitudes.\cite{solenov-gates} The remaining {\it five} free parameters can be set arbitrarily as soon as Eq.~(\ref{eq:romb:ABC}) produce the desired values for $|A|$, $|B|$, and $|C|$. Note that the reversed transformation is also possible. We find,
\begin{eqnarray}\label{eq:romb:bc}
&&\left\{\begin{array}{rcl}
b&=& \frac{d A^* C^* + a^* A B}{A^2}
\\
c&=& \frac{-a A^* C^* + d^* A B}{A^2}
\end{array}\right.
\\\label{eq:romb:ad}
&&|a|^2 + |d|^2 = A^2
\end{eqnarray}
where either $a$ and $\arg d$ or $d$ and $\arg a$ can be chosen arbitrarily, and the values of $|d|$ or $|a|$ (respectively) are found from Eq.~(\ref{eq:romb:ad}).

\section{Six-segment symmetric loop insertion or two-segment branch}\label{sec:2s}

\begin{figure}\begin{center}
\includegraphics[width=0.9\textwidth]{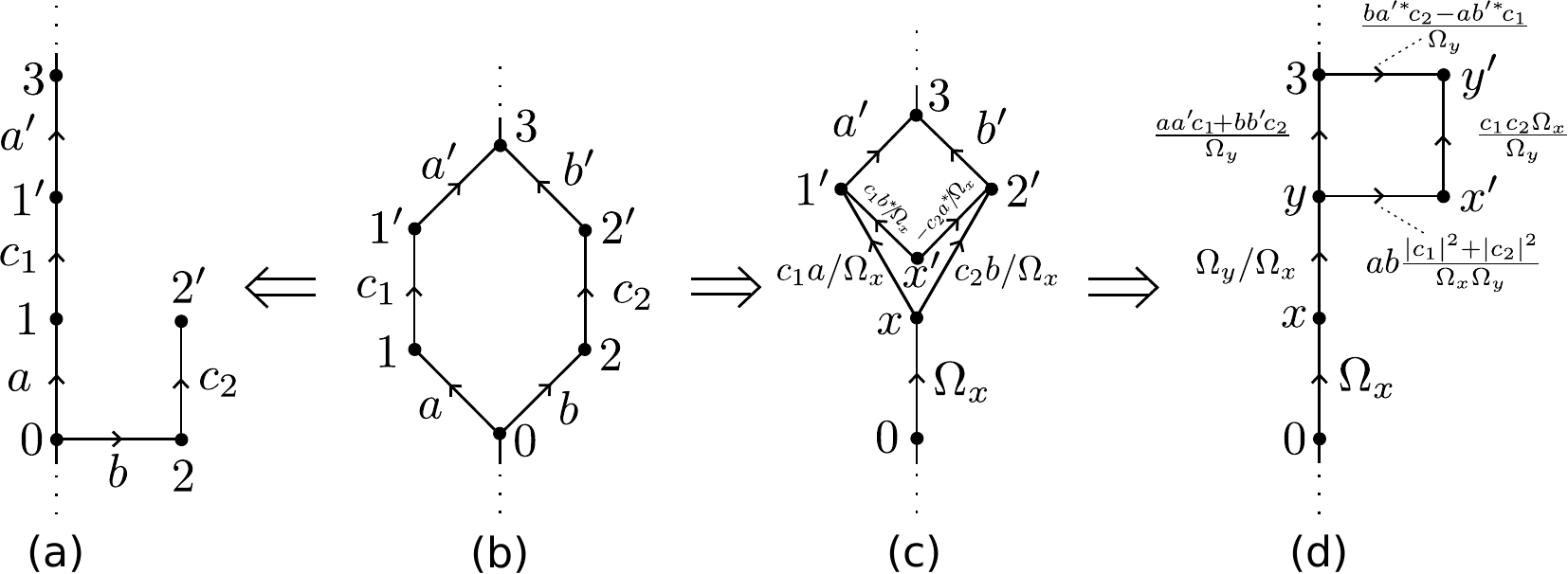}
\caption{\label{fig:diamond}
The transformation between (b) six-segment loop breaking the linear chain graph and a four-segment edge-sharing loop (d). (a) A linear chain with a two-segment branch as a special case of (b). (c) An intermediate graph. Graphic notations are the same as in Fig.~\ref{fig:one-branch}.
}
\end{center}\end{figure}

In this section we derive a transformation that simplifies six-segment symmetric loop insertion shown in Fig.~\ref{fig:diamond}(b), and (as a special case with $b'=0$) a two-segment branch shown in Fig.~\ref{fig:diamond}(a). These graphs, which occur, e.g., in three-qubit gate designs,\cite{solenov-3qb,solenov-gates} can be transformed into an edge-shared four-segment loop shown in Fig.~\ref{fig:diamond}(d). The latter can be further simplified into an linear chain graph as demonstrated in Sec.~\ref{sec:4l}. In all cases continuous time quantum walks through these graphs remain the same, and the graphs [in Figs.~\ref{fig:diamond}(a)-(d) and the linear chain] can be considered topologically equivalent in this sense.

The Hamiltonian corresponding to the subgraphs shown in Fig.~\ref{fig:diamond}(a) and (b) is
\begin{eqnarray}\label{eq:2s:H03}
H_{03} = a \v{0}\iv{1} + b \v{0}\iv{2} 
+ c_1 \v{1}\iv{1'} + c_2 \v{2}\iv{2'}
+ a' \v{1'}\iv{3} + b' \v{2'}\iv{3}
+ h.c.
\end{eqnarray}
The full Hamiltonian $H\{H_{03}\}$ has the form of Eq.~(\ref{eq:1s:H}). As before, the first two terms can be reduced by rotating the basis to introduce new orthogonal states
\begin{eqnarray}\label{eq:2s:xxp}
\left\{\begin{array}{rcl}
\v{x} &=& \frac{a^*\v{1} + b^*\v{2}}{\Omega_x}
\\
\v{x'} &=& \frac{b\v{1} - a\v{2}}{\Omega_x}
\end{array}\right.
\\\label{eq:2s:Omega_x}
\Omega_x = \sqrt{|a|^2 + |b|^2}
\end{eqnarray}
This procedure transforms Eq.~(\ref{eq:2s:H03}) into
\begin{eqnarray}\nonumber
H_{03} &=& \Omega_x \v{0}\iv{x} 
+ \v{x}\frac{ac_1\iv{1'} + bc_2\iv{2'}}{\Omega_x} 
+ \v{x'}\frac{b^*c_1\iv{1'} - a^*c_2\iv{2'}}{\Omega_x} 
\\\label{eq:2s:H03aux}
&&+ a' \v{1'}\iv{3} + b' \v{2'}\iv{3}
+ h.c.
\end{eqnarray}
which corresponds to the graph shown in Fig.~\ref{fig:diamond}(c). Further binary rotation involving states $\v{1'}$ and $\v{2'}$
\begin{eqnarray}\label{eq:2s:yyp}
\left\{\begin{array}{rcl}
\v{y} &=& \frac{a^*c_1^*\v{1} + b^*c_2^*\v{2}}{\Omega_y}
\\
\v{y'} &=& \frac{bc_2\v{1} - ac_1\v{2}}{\Omega_x}
\end{array}\right.
\\\label{eq:2s:Omega_y}
\Omega_y = \sqrt{|ac_1|^2 + |bc_2|^2}
\end{eqnarray}
simplifies the second term in Eq.~(\ref{eq:2s:H03aux}), yielding
\begin{eqnarray}\nonumber
&&H_{03} = \Omega_x \v{0}\iv{x} 
+ \frac{\Omega_y}{\Omega_x}\v{x}\iv{y} 
+ ab\frac{|c_1|^2-|c_2|^2}{\Omega_x\Omega_y}\v{y}\iv{x'}
\\\label{eq:2s:H03new}
&&+ \frac{c_1c_2\Omega_x}{\Omega_y}\v{x'}\iv{y'}
+ \frac{aa'c_1+bb'c_2}{\Omega_y}\v{y}\iv{3}
+ \frac{ba'^*c_2-ab'^*c_1^*}{\Omega_y}\v{3}\iv{y'}
+ h.c.
\end{eqnarray}
The Eq.~(\ref{eq:2s:H03new}) is the adjacency matrix corresponding to an edge-sharing four-segment loop graph shown in Fig.~\ref{fig:diamond}(d). The latter can be further simplified to produce linear chain graph, as explained in Sec.~\ref{sec:4l}, if condition (\ref{eq:4l:cond}) is satisfied. Comparing the last four terms of Eq.~(\ref{eq:2s:H03new}) with Eq.~(\ref{eq:4l:H14:new}) we can rewrite condition (\ref{eq:4l:cond}) as
\begin{eqnarray}\label{eq:2s:cond}
\frac{ba'^*c_2 - ab'^*c_1}{a^*a'^*c_1^*+b^*b'^*c_2^*}
=
\frac{c_1c_2}{a^*b^*}
\frac{|a|^2+|b|^2}{|c_1|^2+|c_2|^2}
\end{eqnarray}
In the special case when $b'=0$ we obtain the transformation procedure for a two-segment branch graph shown in Fig.~\ref{fig:diamond}(a). In this case condition~(\ref{eq:2s:cond}) simplifies to
\begin{eqnarray}\label{eq:2s:cond-2s}
\frac{|b|^2}{|c_1|^2}
=
\frac{|a|^{2}+|b|^2}{|c_1|^2+|c_2|^2}
\end{eqnarray}
This is only possible if
\begin{eqnarray}\label{eq:2s:cond-2s-final}
|c_2|^2|b|^2
=
|a|^2|c_1|^2
\end{eqnarray}
In this case a graph with a two-segment branch [see Fig.~\ref{fig:diamond}(a)] simplifies to a linear chain graph via consecutive rotations discussed here and in Secs.~\ref{sec:4l} and \ref{sec:1s}.

\section{General approach to simplifying quantum walks}\label{sec:Simp}

In general, a Hamiltonian $H$ can be mapped to a linear chain graph of length $n$ with hopping coefficients 
$\Omega_{1}, \Omega_{2},...\Omega_{n-1}$ by finding a basis $\beta=\left\{ \v{x_{1}},\v{x_{2}},..., \v{x_{n}}\right\}$ such that 
the matrix representation of $H$ with respect to $\beta$ is the adjacency matrix of a linear chain graph, which 
is of the form
\begin{eqnarray}\label{eq:AA:Line} 
[H]_{\beta} = \Omega_{1}\v{x_{2}}\iv{x_{1}} + \Omega_{2}\v{x_{3}}\iv{x_{2}}+...+\Omega_{n-1}\v{x_{n}}\iv{x_{n-1}} +h.c.
\end{eqnarray}
Note that the phases of $\Omega_{i}$ can always be lumped into the definition of $\v{x_{i+1}}$, so that $\Omega_{i}$ can be 
chosen to be real. The two endpoint states, $\v{x_{1}}$ 
and $\v{x_{n}}$, are each mapped to a single basis state by $H$. All other basis states $\v{x_{i}}$ 
are sent to a superposition of two basis states, $\v{x_{i-1}} $ and $\v{x_{i+1}} $. The basis $\beta$ 
must satisfy the following equations
\begin{eqnarray}\label{eq:AA:ChainCond}
\left\{\begin{array}{rcl}
 H\v{x_{1}} &=& \Omega_{1}\v{x_{2}} \\
 H\v{x_{i}} &=& \Omega_{i-1}^{*}\v{x_{i-1}}+\Omega_{i}\v{x_{i+1}} \\
 H\v{x_{n}} &=& \Omega_{n-1}^{*}\v{x_{n-1}}
 \end{array}\right.
\end{eqnarray}
for all integers $2\leq i \leq n-1$. A new basis, $\beta$ can be constructed using Eq. (\ref{eq:AA:ChainCond}) and 
Gram-Schmidt orthogonalization.

The method used in the previous sections can be generalized to map graphs to graphs consisting of one, or, in some cases, more linear chains (under certain conditions).
 When the initial graph is bipartite, the resulting linearized 
adjacency matrix does not have diagonal entries. Otherwise, diagonal entries can be generated during the linearization 
procedure. Therefore, bipartite graphs are mapped to bipartite graphs. A bipartite graph is a graph whose vertices can be partitioned into 
two subsets, such that no edge connects two vertices in the same subset. Many potentially useful graphs, 
such as n-cube graphs, are bipartite. 

The first endpoint of the linear chain can be selected as any linear combination of states 
within the same bipartite subset. Often, however, it is important to choose the initial basis state 
to be one of the original basis states. Specifically, when constructing quantum gates, one typically 
starts from one of the computational basis states.\cite{solenov-gates}
Since $ \v{x_{1}} $ is an endpoint, it must be mapped to a single basis state by the Hamiltonian.
Therefore, the next state in the chain, $ \v{x_{2}} $, is defined as
\begin{eqnarray}\label{eq:AA:X2}
\v{x_{2}} &=& \frac{H\v{x_{1}}}{\sqrt{\iv{x_{1}}H^2 \v{x_{1}}}} \\
 \Omega_{1} &=& \iv{x_{2}}H\v{x_{1}}
\end{eqnarray}

The Hamiltonian then maps the state $ \v{x_{2}} $ to some linear combination of 
the previous state, $ \v{x_{1}} $, and the next state, $ \v{x_{3}} $,
\begin{eqnarray}\label{eq:AA:HX2}
 H\v{x_{2}} = \Omega_{1}\conj \v{x_{1}} + \Omega_{2}\v{x_{3}} 
\end{eqnarray}
Where $\Omega_{1}$ and $\Omega_{2}$ are hopping amplitudes of the new graph. In order to determine what $ \v{x_{3}} $ is, the projection of $ H\v{x_{2}} $ onto $ \v{x_{1}} $
is subtracted from $ H\v{x_{2}} $
\begin{eqnarray}\label{eq:AA:ProjSub}
H\v{x_{2}} -  \iv{x_{1}}H\v{x_{2}} \v{x_{1}} &=& \Omega_{2} \v{x_{3}} 
\end{eqnarray}
From this equation, it follows that
\begin{eqnarray}\label{eq:AA:X3}
 \v{x_{3}} &=& \frac{ H\v{x_{2}} - \Omega_{1}^{*}\v{x_{1}}}{\sqrt{\iv{x_{2}}H^{2}\v{x_{2}}-|\Omega_{1}|^{2}}} \\
 \Omega_{2} &=& \iv{x_{3}}H\v{x_{2}} 
\end{eqnarray}
 This process can be repeated recursively; the next state
in the linear chain is defined by applying the Hamiltonian to the previous link of in chain, and
then subtracting off the projection onto the previous state. In general, one iteration step is given by 
\begin{eqnarray}\label{eq:AA:GenX}
 \v{x_{k+1}} &=& \frac{H\v{x_{k}} - \Omega_{k-1}^{*}\v{x_{k-1}}}{\sqrt{\iv{x_{k}}H^{2}\v{x_{k}}-|\Omega_{k-1}|^{2}}} \\
 \Omega_{k} &=& \iv{x_{k+1}}H\v{x_{k}}
 \end{eqnarray}
When the graph is bipartite, this process will not create any self-loops 
(diagonal entries of the adjacency matrix), as demonstrated in \ref{app:AA:OrthoProof}. This procedure continues until it generates a state 
$ \v{x_{n}} $ such that it is mapped only to the previous state in the chain by the Hamiltonian, see Eq. (\ref{eq:AA:ChainCond}).
The state $ \v{x_{n}} $ becomes the other endpoint of
the linear chain. If the linear chain has as many vertices as the original graph, this completes 
the new basis $\beta$. If it does not, then a new state that is orthogonal to each state in $\beta$ 
is chosen to be an endpoint of the second linear chain. From this endpoint, another linear chain is created using the same method. 
This process can be repeated until $\beta$ spans the entire space.

It is instructive to further examine the case where the starting endpoint of $\beta$ is a state in the 
original basis. This is the case, for instance, when one is constructing quantum gates based on quantum walks.\cite{solenov-gates}
To simplify derivations, we can introduce two bases: an auxiliary basis that is not normalized, $\gamma = \left\{ \v{y_{1}},\v{y_{2}},...,\v{y_{n}} \right\}$,
and a normalized basis, $\beta = \left\{\v{x_{1}},\v{x_{2}},...,\v{x_{n}}\right\}$. The latter is the basis 
that will be used for the linear chain, therefore it must satisfy Eq. (\ref{eq:AA:ChainCond})
The non-normalized basis can be constructed by choosing $\v{y_{1}}=\v{x_{1}} $ to be 
a state in the original basis, and defining each subsequent basis state
\begin{eqnarray}\label{eq:AA:YBas}
 \v{y_{2}} &=& H\v{y_{1}} \\ \label{eq:AA:yp1}
 \v{y_{i+1}} &=& H\v{y_{i}}-\frac{\Omega_{i-1}^{*}N_{i}}{N_{i-1}} \v{y_{i-1}}
\end{eqnarray}
Where $N_{j} = \sqrt{\iv{y_{j}}\v{y_{j}}}$ is the magnitude of $\v{y_{j}}$, and 
$\Omega_{i} = \frac{\iv{y_{i+1}}H\v{y_{i}}}{N_{i+1}N_{i}}$. Note that $N_{i}$, as well as the state 
$\v{y_{i}}$, both have units of energy to the power of $i-1$. Note also that if we solve Eq. (\ref{eq:AA:yp1}) 
for $H\v{y_{i}}$ and use it to compute $\iv{x_{i+1}}H\v{x_{i}} $, we obtain
\begin{eqnarray}\label{eq:AA:LDef}
 \Omega_{j} = \frac{N_{j+1}}{N_{j}}
\end{eqnarray}
New basis states are generated until the second endpoint 
is found, i.e. $\v{y_{n}}$.
\begin{eqnarray}\label{eq:AA:YnDef}
 H\v{y_{n}} = \frac{\Omega_{i-1}^{*}N_{n}}{N_{n-1}}\v{y_{n-1}}
\end{eqnarray}
Each $\v{x_{j}}$ in $\beta$ is then defined by normalizing $\v{y_{j}}$.
\begin{eqnarray}\label{eq:AA:XDef}
 \v{x_{j}} = \frac{\v{y_{j}}}{N_{j}}
\end{eqnarray}

As before, if the obtained states do not span the space of the original graph, then another state is chosen
to start a new linear chain. 
\begin{figure}\begin{center}
\includegraphics[width=0.65\textwidth]{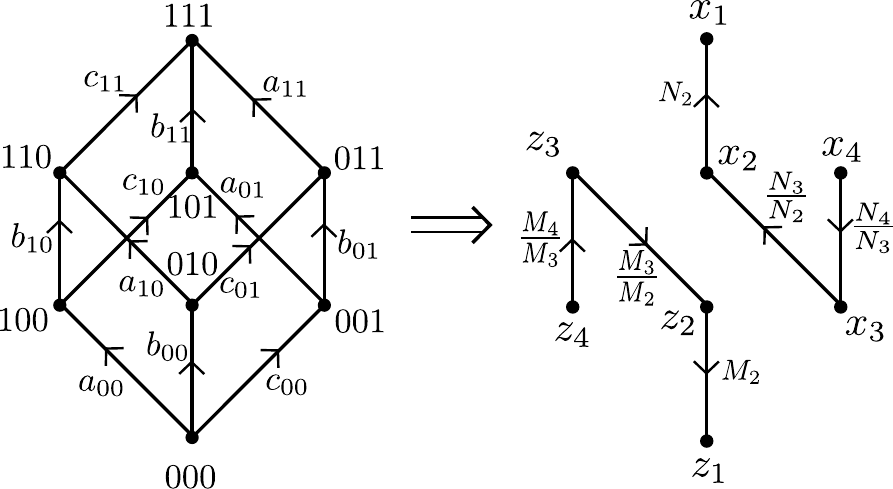}
\caption{\label{CubeSplit}
The transformation between three-dimensional cube and two disjoint linear chains. The graphic notation is the 
same as in Fig. \ref{fig:one-branch}
}
\end{center}\end{figure}

\section{Example: reducing cube graph}\label{sec:ExCube}

In this section we derive a transformation that maps the cube graph shown in Fig. \ref{CubeSplit} to a graph of two disjoint linear chains 
of four nodes each, to illustrate the procedure introduced in section \ref{sec:Simp}.

\subsection{Notation}

The vertices of the cube can be thought of as corners of a unit cube in the first 
octant of $\mathbb{R}^{3}$. The state $\v{ijk}$ corresponds to the corner with Cartesian Coordinates $(i,j,k)$,
where $i,j,k$ are 0 or 1. This notation is also natural when discussing multi-qubit quantum gates 
in systems with multiple auxilliary states.\cite{solenov-gates} Complex hopping coefficients on edges parallel to the x-axis are denoted $a_{jk}$, 
coefficients parallel to the y-axis are denoted $b_{ik}$, and coefficients parallel to the z-axis are 
denoted $c_{ij}$, with the subscripts denoting the coordinates that are unchanged by the transition. 
These coefficients are associated with $1\rightarrow0$ transitions. The coefficients associated with $0\rightarrow1$ 
transitions are complex conjugates of the corresponding $1\rightarrow0$ transitions.
The Hamiltonian corresponding to this graph is 
\begin{eqnarray}\label{eq:CG:CubeHam} \nonumber
 H &=& \v{000} \left( a_{00}\iv{100}+b_{00}\iv{010}+c_{00}\iv{001}\right)+\left( b_{01}\v{001}+c_{01}\v{010} \right)\iv{011} \\ \nonumber
 &+&\left( a_{01}\v{001}+c_{10}\v{001} \right)\iv{101} + \left( a_{10}\v{001}+b_{10}\v{010} \right)\iv{110} \\
 &+&\left(a_{11}\v{011}+b_{11}\v{101}+c_{11}\v{110} \right)\iv{111}+h.c.
\end{eqnarray}
In the next subsections we will focus on a special case where the coefficients on parallel edges differ only by phases, i.e., all coefficients with the same letter will be set to have the same magnitude. This magnitude is denoted by a letter without a subscript.
\begin{eqnarray}\label{eq:CG:Mag}
\left\{\begin{array}{rcl}
 a &=& \left| a_{00} \right| = \left| a_{01} \right| = \left| a_{10} \right| = \left| a_{11} \right| \\
 b &=& \left| b_{00} \right| = \left| b_{01} \right| = \left| b_{10} \right| = \left| b_{11} \right| \\
 c &=& \left| c_{00} \right| = \left| c_{01} \right| = \left| c_{10} \right| = \left| c_{11} \right|
\end{array}\right.
 \end{eqnarray}
Furthermore, we will search for a solution for which the cube graph splits into
two symmetrical chains of four nodes each. This imposes two conditions,
\begin{eqnarray}\label{eq:CG:Cond1}
a_{01}b_{10}c_{01} + a_{10}b_{01}c_{10} = 0 \\ \label{eq:CG:Cond2}
a_{00}\left(b_{11}c_{10} + b_{10}c_{11}\right) +  b_{00}\left(a_{11}c_{01} + a_{10}c_{11}\right) + c_{00}\left(a_{11}b_{01} + a_{01}b_{11}\right) = 0
\end{eqnarray}
derived in the next section.
Note that these conditions depend only on the phases of the coefficients, not the magnitudes, if condition (\ref{eq:CG:Mag}) is assumed.
The complex phases of hopping amplitudes will be denoted with the corresponding Greek letter with the same subscript,
e.g., $a_{01} = ae^{i\alpha_{01}}$. Note also that conditions (\ref{eq:CG:Cond1}) are, in fact, more general, and do not require assumption (\ref{eq:CG:Mag}), i.e.
amplitudes can vary.

\subsection{Cube splitting solution}
In this subsection, we derive a transformation that simplifies a cube graph under conditions stated in 
Eq. (\ref{eq:CG:Cond1}) and Eq. (\ref{eq:CG:Cond2}) into two independent linear chains. These linear 
chains have four nodes each, which makes quantum walks on this graph solveable analytically.\cite{solenov-gates}

The first element of the new basis $\beta$ is chosen to be $\v{111}$
\begin{eqnarray}\label{eq:CG:X1}
\v{x_{1}} = \v{111}
\end{eqnarray}
Acting with the Hamiltonian on $\v{x_{1}}$ gives the state
\begin{eqnarray}\label{eq:CG:Y2}
\v{y_{2}} = H\v{x_{1}} = a_{11}\v{011}+b_{11}\v{101}+c_{11}\v{110}
\end{eqnarray}
which is divided by a normalizing factor
\begin{eqnarray}\label{eq:CG:N2}
N_{2} = \sqrt{a^{2}+b^{2}+c^{2}}
\end{eqnarray}
to obtain $\v{x_{2}}$ Now, we act with the Hamiltonian on $\v{y_{2}}$ and discard all $\v{x_{1}}\iv{y_{2}}$ terms, which gives
\begin{eqnarray}\label{eq:CG:Y3} \nonumber
\v{y_{3}} &=& \left(b_{11}c_{10}+c_{11}b_{10}\right)\v{100} 
 +\left(a_{11}c_{01}+c_{11}a_{10}\right)\v{010} \\
 &+&\left(a_{11}b_{01}+b_{11}a_{01}\right)\v{001}
\end{eqnarray}
This must be divided by the normalizing factor
\begin{eqnarray}\label{eq:CG:N3} 
N_{3} = 2\sqrt{a^{2}b^{2}\cos^{2}\frac{\phi_{1}}{2}+a^{2}c^{2}\cos^{2}\frac{\phi_{2}}{2}+a^{2}+b{2}c^{2}\cos^{2}\frac{\phi_{3}}{2}}
\end{eqnarray}
to produce $\v{x_{3}}$.
Here the $\phi$ phases are defined as
\begin{eqnarray}\label{eq:CG:PhiDef}
\left\{\begin{array}{rcl}
 \phi_{1} &=& \alpha_{11}-\beta_{11}-\alpha_{01}+\beta_{01} \\
 \phi_{2} &=& \gamma_{11}-\alpha_{11}-\gamma_{01}+\alpha_{10} \\
 \phi_{3} &=& \beta_{11}-\gamma_{11}-\beta_{10}+\gamma_{10}
 \end{array}\right.
\end{eqnarray}
A detailed derivation of $N_{3}$ can be found in \ref{app:N3Der}.
Because the phase of $\Omega_{2}$ can be absorbed into $\v{x_{3}}$, 
$\Omega_{2}$ can be defined as $N_{3}/N_{2}$.

In general, $H$ will map $\v{y_{3}}$ to a state  $\sim\v{000}$.
To ensure that the second linear chain can be created with $\v{000}$ as an endpoint, we must have 
$\iv{000}H\v{y_{3}}=0$. The only term in the Hamiltonian (\ref{eq:CG:CubeHam}) contributing to this is the first term, which, 
in combination with Eq. (\ref{eq:CG:Y3}), gives the second cube splitting condition, Eq. (\ref{eq:CG:Cond2}). 

The next, and final, basis state of the first linear chain, $\v{y_{4}}$, is found by applying the Hamiltonian to $\v{y_{3}}$ and subtracting off the 
projection onto $\v{x_{2}}$. 
\begin{eqnarray}\label{eq:CG:y4}
\v{y_{4}} &=& \left[a_{11}\left(b^{2}+c^{2}-\frac{N_{3}^{2}}{N_{2}^{2}}\right) + b_{11}a_{01}b_{01}\conj + c_{11}a_{10}c_{01}\conj \right]\v{011}\ \\ \nonumber
 &+&\left[a_{11}b_{01}a_{01}\conj + b_{11}\left(a^{2}+c^{2}-\frac{N_{3}^{2}}{N_{2}^{2}}\right) + c_{11}b_{10}c_{10}\conj \right]\v{101}\\
 &+& \left[a_{11}c_{01}a_{10}\conj + b_{11}c_{10}b_{10}\conj + c_{11}\left(a^{2}+b^{2}-\frac{N_{3}^{2}}{N_{2}^{2}}\right) \right]\v{110}
\end{eqnarray}
This state is divided by the normalizing factor
\begin{eqnarray}\label{eq:CG:N4}
N_{4} = \frac{N_{3}}{N_{2}}\sqrt{ N_{2}^{4}-N_{3}^{2}}
\end{eqnarray}
to get $\v{x_{4}}$.
For a full derivation of $\v{y_{4}}$ and $N_{4}$, see \ref{app:X4Der}. 

In order to guarantee that $\v{y_{4}}$ is the endpoint of the chain, we must have
\begin{eqnarray}\label{eq:AppEnd:HY4} \nonumber
&&H\v{y_{4}}=\\ \nonumber
&&+ \left[a_{11}\left(a_{01}^{*}b_{01}c_{10}+a_{10}^{*}b_{10}c_{01}\right) +\left(b_{11}c_{10}+c_{11}b_{10}\right) \left( a^{2}+b^{2}+c^{2}-\frac{N_{3}^{2}}{N_{2}^{2}}\right) \right]\v{100} \\ \nonumber 
&&+\left[b_{11}\left(a_{10}b_{10}^{*}c_{10}+a_{01}b_{01}^{*}c_{01}\right) +\left(a_{11}c_{01}+c_{11}a_{10}\right) \left( a^{2}+b^{2}+c^{2}-\frac{N_{3}^{2}}{N_{2}^{2}}\right) \right]\v{010} \\ \nonumber
&&+\left[c_{11}\left(a_{10}b_{01}c_{01}^{*}+a_{01}b_{10}c_{10}^{*}\right) +\left(a_{11}b_{01}+b_{11}a_{01}\right) \left( a^{2}+b^{2}+c^{2}-\frac{N_{3}^{2}}{N_{2}^{2}}\right) \right]\v{010}
\end{eqnarray}
be a multiple of state $\v{y_{3}}$.
This occurs when the first term of each line vanishes, which happens when the first cube splitting condition, 
Eq. (\ref{eq:CG:Cond1}), is satisfied.

The second linear chain that contains $\v{000}$ is constructed similarly. Because the system is (symbolically) symmetric 
with respect to inverting each index, we can obtain the second linear chain by inverting indices of every state 
of the first linear chain. Because all $1\rightarrow0$ transitions become $0\rightarrow1$ transitions, 
and vice versa, all coefficients must also be complex conjugated. 
\begin{eqnarray}\label{eq:CG:ZDef}
\left\{\begin{array}{rcl}
\v{z_{1}} &=& \v{000} \\
\v{z_{2}} &=& \frac{a_{00}^{*}\v{100}+b_{00}^{*}\v{010}+c_{00}^{*}\v{001}}{M_{2}} \\ 
\v{z_{3}} &=& \frac{1}{M_{3}}[\left(b_{00}^{*}c_{01}^{*}+c_{00}^{*}b_{01}^{*}\right)\v{011} \\ 
 &+&\left(a_{00}^{*}c_{10}^{*}+c_{00}^{*}a_{01}^{*}\right)\v{101}\\
 &+&\left(a_{00}^{*}b_{10}^{*}+b_{00}^{*}a_{10}^{*}\right)\v{110}] \\ 
 \v{z_{4}} &=&  \frac{1}{M_{4}}\left[a_{00}^{*}\left(b^{2}+c^{2}-\frac{M_{3}^{2}}{M_{2}^{2}}\right) + b_{00}^{*}a_{10}^{*}b_{10} + c_{00}^{*}a_{01}^{*}c_{10} \right]\v{100}\ \\ 
 &+&\left[a_{00}^{*}b_{10}^{*}a_{10} + b_{00}^{*}\left(a^{2}+c^{2}-\frac{M_{3}^{2}}{M_{2}^{2}}\right) + c_{00}^{*}b_{01}^{*}c_{01} \right]\v{101}\\
 &+& \left[a_{00}^{*}c_{10}^{*}a_{01} + b_{00}^{*}c_{01}^{*}b_{01} + c_{00}^{*}\left(a^{2}+b^{2}-\frac{M_{3}^{2}}{M_{2}^{2}}\right) \right]\v{110}
\end{array}\right.
 \end{eqnarray}
 The hopping amplitudes of the second chain, $\Omega_{i}'$, are given by 
 \begin{eqnarray}
  \Omega_{i}' = \frac{M_{i+1}}{M_{i}}
 \end{eqnarray}
where the normalizing constants $M_{1}$, $M_{2}$, and $M_{3}$ are
\begin{eqnarray}\label{eq:CG:MDef}
\left\{\begin{array}{rcl}
M_{2} &=& \sqrt{a^{2}+b^{2}+c^{2}} \\
M_{3} &=& 2\sqrt{a^{2}b^{2}\cos^{2}\frac{\theta_{1}}{2}+a^{2}c^{2}\cos^{2}\frac{\theta_{2}}{2}+b^{2}c^{2}\cos^{2}\frac{\theta_{3}}{2}} \\
M_{4} &=& \frac{M_{3}}{M_{2}}\sqrt{ M_{2}^{4}-M_{3}^{2}}
\end{array}\right.
\end{eqnarray}
and the $\theta_{i}$ phases are defined as
\begin{eqnarray}\label{eq:CG:ThetaDef}
\left\{\begin{array}{rcl}
 \theta_{1} &=& \alpha_{00}-\beta_{00}-\alpha_{10}+\beta_{10} \\
 \theta_{2} &=& \gamma_{00}-\alpha_{00}-\gamma_{10}+\alpha_{01} \\
 \theta_{3} &=& \beta_{00}-\gamma_{00}-\beta_{01}+\gamma_{01}
 \end{array}\right.
\end{eqnarray}
Note that the cube splitting conditions 
(\ref{eq:CG:Cond1}) and (\ref{eq:CG:Cond2}) are unchanged by bit-flipping. Thus, they also ensure that 
the chain starting at $\v{000}$ terminates at the fourth node as well.
All the newfound states form an orthogonal basis.
The Hamiltonian takes the form 
\begin{eqnarray}\label{eq:CG:LineHam} \nonumber 
H &=& N_{2}\v{x_{2}}\iv{x_{1}} +\frac{N_{3}}{N_{2}}\v{x_{3}}\iv{x_{2}}+\frac{N_{4}}{N_{3}}\v{x_{4}}\iv{x_{3}}+h.c \\
&+& M_{2}\v{z_{2}}\iv{z_{1}} +\frac{M_{3}}{M_{2}}\v{z_{3}}\iv{z_{2}}+\frac{M_{4}}{M_{3}}\v{z_{4}}\iv{z_{3}}+h.c
\end{eqnarray}
as shown in Figure \ref{CubeSplit}.

\subsection{Deriving return walk on cube graph}

Once the cube graph has been mapped to two linear chains of four nodes, the solution derived in 
section IV.C of Ref.~\refcite{solenov-gates} can be used to describe continuous time quantum walks in this system.
The complete solution must satisfy four equations
\begin{eqnarray}\label{eq:CG:SYS1}
&&a_{01}b_{10}c_{01} + a_{10}b_{01}c_{10} = 0 \\ \label{eq:CG:SYS12}
&&a_{00}\left(b_{11}c_{10} + b_{10}c_{11}\right) +  b_{00}\left(a_{11}c_{01} + a_{10}c_{11}\right) + c_{00}\left(a_{11}b_{01} + a_{01}b_{11}\right) = 0
\\ \label{eq:CG:SYS13}
&&\abs{\Omega_{1}}^{2}+\abs{\Omega_{2}}^{2}+\abs{\Omega_{3}}^{2} = n^{2} + m^{2}
\\ \label{eq:CG:SYS14}
&&\abs{\Omega_{1}}\abs{\Omega_{3}} = nm
\end{eqnarray}
The first two are the conditions that ensure the cube graph can be
split, and the last two ensure that the quantum evolution operator describing the walk is $U(\tau) = e^{-i\tau H} = \pm1$, i.e.,
the particle returns to the initial state with or without an added phase [see Eq. (59) of Ref.~\refcite{solenov-gates}]. For the purpose of calculating quantum walks, it is convenient to introduce a dimensionless Hamiltonian 
$H' = H\tau/\pi$, and lump $\tau/ \pi$ in with each of the $a$, $b$, $c$ and $\Omega$ coefficients,
making them dimensionless as well. Therefore, from now on, we consider all coefficients to be dimensionless quantities. 
The evolution operator is $U(\tau) = e^{-i \pi H'}$.

In what follows, we will derive a specific solution to return quantum walks as an illustration. 
This solution can have eleven parameters:
real magnitudes, $a$, $b$, and $c$, integers $n$ and $m$ of equal parity, and six phases, $\alpha_{11},
\alpha_{00}, \beta_{11}, \beta_{00}, \gamma_{11},$ and $\gamma_{00}$. The first 
amplitude is $\Omega_{1}=N_{2}$, and the others are found using Eqs. (\ref{eq:AA:LDef}), (\ref{eq:CG:N2}),
(\ref{eq:CG:N3}), and (\ref{eq:CG:N4}). Using these equations with Eq. (\ref{eq:CG:SYS13}) gives 
the following condition on the magnitudes
\begin{eqnarray}\label{eq:CG:SolCond}
 2\left(a^{2}+b^{2}+c^{2}\right) = n^{2}+m^{2}
 \end{eqnarray}
 There is another condition,
 \begin{eqnarray}\label{eq:CG:SolCond2}
 4c\sqrt{a^{2}+b^{2}}\geq \abs{n^{2}-m^{2}}
\end{eqnarray}
which accounts for Eq. (76), and several other assumptions introduced later in this subsection. 
It will be derived by the end of this subsection. 
The remaining six phases are then expressed in terms of these adjustable parameters to ensure that all 
four equations, i.e., Eqs. (\ref{eq:CG:SYS1}), (\ref{eq:CG:SYS12}), (\ref{eq:CG:SYS13}), (\ref{eq:CG:SYS14}), hold.

In order to solve this system, we express Eqs. (\ref{eq:CG:SYS1}), (\ref{eq:CG:SYS12}), (\ref{eq:CG:SYS14})
in terms of $\phi_{1}$, $\phi_{2}$, $\phi_{3}$, defined in Eq. (\ref{eq:CG:PhiDef}).
Recall that each coefficient in (\ref{eq:CG:SYS1}) can be expressed in terms of its magnitude and its phase. 
This gives
\begin{eqnarray}\label{eq:CG:Cond1Phase}
 \alpha_{01}+\beta_{10}+\gamma_{01} = \alpha_{10}+\beta_{01}+\gamma_{10} + (2k+1)\pi 
 \end{eqnarray}
where $ k $ is any integer number.
This can be written in terms of the $ \phi $ phases by pulling all phases to one side
\begin{eqnarray}\label{eq:CG:Cond1'}
  \phi_{1}+\phi_{2}+\phi_{3} = (2k+1)\pi
\end{eqnarray}
Plugging the transition coefficients $\Omega_{i}$, derived in the previous section, into Eq. (\ref{eq:CG:SYS14}) gives
\begin{eqnarray}\nonumber
 \sqrt{\left(a^{2}+b^{2}+c^{2} \right)^{2}-4\left(a^{2}b^{2}\cos^{2}\frac{\phi_{1}}{2}+a^{2}c^{2}\cos^{2}\frac{\phi_{2}}{2}+b^{2}c^{2}\cos^{2}\frac{\phi_{3}}{2}\right)} &=& nm \\ \nonumber
 \left(\frac{n^{2}+m^{2}}{2}\right)^{2}-4\left(a^{2}b^{2}\cos^{2}\frac{\phi_{1}}{2}+a^{2}c^{2}\cos^{2}\frac{\phi_{2}}{2}+b^{2}c^{2}\cos^{2}\frac{\phi_{3}}{2}\right) &=& n^{2}m^{2}
\end{eqnarray}
which can be rearranged to 
\begin{eqnarray}\label{eq:CG:PhiEq2}
 a^{2}b^{2}\cos^{2}\frac{\phi_{1}}{2}+a^{2}c^{2}\cos^{2}\frac{\phi_{2}}{2}+b^{2}c^{2}\cos^{2}\frac{\phi_{3}}{2}&=& \left(\frac{n^{2}-m^{2}}{4}\right)^{2}
\end{eqnarray}
In order to express Eq. (\ref{eq:CG:SYS12}) in terms of the 
$\phi$ phases, we first divide it by $abc$ to obtain
\begin{eqnarray}\label{eq:CG:eq2phase} \nonumber
& &e^{i\alpha_{00}}\left[e^{i\left(\beta_{11}+\gamma_{10}\right)} + e^{i\left(\beta_{10}+\gamma_{11}\right)} \right] \\ \nonumber 
 &+&e^{i{b_{00}}}\left[\expi{\alpha_{11}+\gamma_{01}} + \expi{\alpha_{10}+\gamma_{11}}\right]  \\ \label{eq:CG:PhEq2}
&+&e^{ic_{00}}\left[\expi{\alpha_{11}+\beta_{01}} + \expi{\alpha_{01}+\beta_{11}}\right] = 0
\end{eqnarray}
We can then take the definitions of $\phi_{1}, \phi_{2},$ and $\phi_{3}$, given in 
Eq. (\ref{eq:CG:PhiDef}), and rearrange them to obtain
\begin{eqnarray}\label{eq:CG:PhiSub}
\left\{\begin{array}{rcl}
 \alpha_{11}+\beta_{01} = \phi_{01}+\alpha_{01}+\beta_{11} \\
 \alpha_{10}+\gamma_{11}=\phi_{2}+\alpha_{11}+\gamma_{01} \\
 \beta_{11}+\gamma_{10} = \phi_{3}+\beta_{10}+\gamma_{11}
 \end{array}\right.
\end{eqnarray}
Using this system, we can simplify Eq. (\ref{eq:CG:eq2phase}) 
to
\begin{eqnarray}\label{eq:CG:Cos1} \nonumber
 & &\exp\left[i\left({\alpha_{00}+\beta_{10}+\gamma_{11}+\frac{\phi_{3}}{2}}\right)\right]\cos\frac{\phi_{3}}{2} \\ \nonumber
 &+&\exp\left[i\left({\alpha_{11}+\beta_{00}+\gamma_{01}+\frac{\phi_{2}}{2}}\right)\right]\cos\frac{\phi_{2}}{2} \\ 
 &+&\exp\left[i\left({\alpha_{01}+\beta_{11}+\gamma_{00}+\frac{\phi_{1}}{2}}\right)\right]\cos\frac{\phi_{1}}{2} = 0
\end{eqnarray}
In order to choose one specific solution as an example, we assume
\begin{eqnarray}\label{eq:CG:Phase1}
\left\{\begin{array}{rcl}
 \alpha_{01} = -\beta_{11}-\gamma_{00}-\frac{\phi_{1}}{2} \\
 \beta_{10} = -\alpha_{00}-\gamma_{11}-\frac{\phi_{3}}{2} \\
 \gamma_{01} = -\alpha_{11}-\beta_{00}-\frac{\phi_{2}}{2}
 \end{array}\right.
\end{eqnarray}
The remaining phases can be calculated by substituting the above equations into the definitions 
of the $\phi$ phases, Eq. (\ref{eq:CG:PhiDef}). We obtain
\begin{eqnarray}\label{eq:CG:Phase2}
\left\{\begin{array}{rcl}
 \alpha_{10} = -\beta_{00}-\gamma_{11}+\frac{\phi_{2}}{2} \\
 \beta_{01} = -\alpha_{11}-\gamma_{00}+\frac{\phi_{1}}{2} \\
 \gamma_{10} = -\alpha_{00}-\beta_{11}+\frac{\phi_{3}}{2}
 \end{array}\right.
\end{eqnarray}
After this assumption, Eq. (\ref{eq:CG:Cos1}) simplifies to
\begin{eqnarray}\label{eq:CG:PhiEq3}
 \cos\frac{\phi_{1}}{2}+\cos\frac{\phi_{2}}{2}+\cos\frac{\phi_{3}}{2} = 0
\end{eqnarray}
As a result, we obtain a system of equations, i.e., Eqs. (\ref{eq:CG:Cond1'}), (\ref{eq:CG:PhiEq3}), (\ref{eq:CG:PhiEq2}),
with only three unknowns, $\phi_{1}$, $\phi_{2}$, and $\phi_{3}$.

So far we have reduced a system of four equations with six unknowns into a system of three equations 
with three unknowns. This has been accomplished by assuming relation (\ref{eq:CG:Phase1})
to simplify Eq. (\ref{eq:CG:Cos1}). In principle, other conditions can be formulated to satisfy 
Eq. (\ref{eq:CG:Cos1}). In what follows, we will further reduce this system by introducing two more 
assumptions in order to narrow down the solution.

If we introduce new variables,
\begin{eqnarray}\label{eq:CG:xyzdef}
\left\{\begin{array}{rcl}
x = \cos\frac{\phi_{1}}{2} \\
y = \cos\frac{\phi_{2}}{2} \\
z = \cos\frac{\phi_{3}}{2}
 \end{array}\right.
\end{eqnarray}
we can rewrite Eqs. (\ref{eq:CG:Cond1'}), (\ref{eq:CG:PhiEq3}), (\ref{eq:CG:PhiEq2}) as
\begin{eqnarray}\label{eq:CG:xyzSys} 
\left\{\begin{array}{rcl}
 xyz = x\sqrt{1-y^{2}}\sqrt{1-z^{2}}+y\sqrt{1-x^{2}}\sqrt{1-z^{2}} +z\sqrt{1-x^{2}}\sqrt{1-y^{2}} \\
 x+y+z = 0 \\ 
 a^{2}b^{2}x^{2}+a^{2}c^{2}y^{2}+b^{2}c^{2}z^{2} = \left(\frac{n^{2}-m^{2}}{4} \right)^{2}
\end{array}\right.
\end{eqnarray}
We now assume $\cos(\phi_{1}/2) = x=0$ and $\cos(\phi_{3}/2)= z=-y=\cos(\phi_{2}/2)$, which will narrow our solution further by making the first two 
equations hold. Finally, we have
\begin{eqnarray}\label{eq:CG:xyz1eq}
\phi_{1} = \pi,\quad \quad \phi_{2} = 2\pi - \phi_{3},\quad \quad \cos\phi_{2} = \frac{n^{2}-m^{2}}{4c\sqrt{a^{2}+b^{2}}}
\end{eqnarray}
The last equation in solution (\ref{eq:CG:xyz1eq}) defines inequality condition (\ref{eq:CG:SolCond2}).
These $\phi$ phases can be used in Eqs. (\ref{eq:CG:Phase1}) and (\ref{eq:CG:Phase2}) to find the remaining 
phases $\alpha_{01}, \alpha_{10}, \beta_{01}, \beta_{10}, \gamma_{01}, \gamma_{10}$. If $n$ and $m$ are different 
even integers and  $a$, $b$, 
$c$, satisfy the conditions (\ref{eq:CG:SolCond}) and (\ref{eq:CG:SolCond2}), we produce a quantum walk with the trivial evolution 
operator $U = 1$. On the other hand, if we set $n$ and $m$ to be non-equal odd integers, we obtain a quantum 
walk with the evolution operator $U = -1$, i.e., if the walk begins at $\v{111}$, it will be returning 
to that node with a nontrivial phase of $\pi$.

\section*{Acknowledgments}

This research was supported by SLU.

\appendix

\section{Proof of orthogonality}\label{app:AA:OrthoProof}

In order to show that $\beta$ is an orthogonal basis, it suffices to show 
that each new state is orthogonal to all previous states. That is, $ \iv{x_{i}} \v{x_{k}} = 0 $ 
for all $ i $ between 1 and $ k - 1$. This is proven by complete induction on $k$.

\subsection{Parity-orthogonality lemma}

If $ i $ and $ j $ have different parity, then $ \v{x_{j}} $ and $ \v{x_{i}} $ are orthogonal states,
that is, 
\begin{eqnarray}\label{eq:AppA:POL}
 i\not\equiv j \mod 2 \Rightarrow \iv{x_{i}} \v{x_{j}} = 0
\end{eqnarray}
This results from the fact that the original graph is bipartite, meaning that the vertices can be 
partitioned into two subsets, A and B, such that no edge connects two vertices in the same subset. 
The state $ \v{x_{1}} $ is chosen to be a linear combination of states in subset A.
Since the Hamiltonian always transitions states between the two subsets, repeatedly applying the 
Hamiltonian to a state in subset A will always produce a state that is entirely contained in one of the 
two subsets. Because all states in the chain are produced by applying the Hamiltonian and subtracting 
off projections, this means that each state in the linear chain lies entirely in A or entirely in B. 
If the Hamiltonian is applied an even number of times, the resulting state is in A. Otherwise, the resulting 
state is in B. Thus, all states with odd subscript are in A, and all states with even subscript are in B. 
If two states have different parity, then they are in different subsets. Therefore, the states must be 
orthogonal.
A corollary of this is that if $ \v{x_{i}} $ is a state that is entirely contained in one of the subsets, 
then $ H'\v{x_{i}}=\Omega_{i}\v{x_{i+1}}+\Omega_{i-1}^{*}\v{x_{i-1}} $ is orthogonal to $ \v{x_{i}} $.

We have the base case $k = 2 $, where $ \iv{x_{1}}\v{x_{2}} = 0 $ because the subscripts have different parity.
Because the formula for $ \v{x_{k+1}} $ includes $ \v{x_{k-1}} $, it is necessary to also consider 
$ k = 3 $ as a base case. We have $ \iv{x_{2}}\v{x_{3}} = 0 $ because the subscripts have different parity. We also see that 
$ \iv{x_{1}}\v{x_{3}} = 0 $ because $ \v{x_{3}} $ is computed by subtracting off the projection onto $ \v{x_{1}} $.
For the inductive hypothesis, assume that $ \iv{x_{i}} \v{x_{k}} = 0 $ for all integers  $ i $ and $ k $ such that 
$ 1 \leq i \leq n $, $ 1 \leq k \leq n $, $ i \neq k $, for some positive integer $ n $.
Now, show that $ \iv{x_{i}} \v{x_{n+1}} = 0 $ for all $1 \leq i \leq n $. 
In the case that $ i = n-1 $, this state $\v{{x_{n+1}}}$ is computed by subtracting off the projection onto ${x_{n-1}}$, 
therefore the two states are orthogonal. In the case that $ i \neq n-1 $, note that 
\begin{eqnarray}\label{eq:AppA:Hdot} 
 \iv{x_{i}}H\v{x_{n}} = \iv{x_{i}} \left( \Omega_{n}\conj \v{x_{n-1}}+ \Omega_{n+1} \v{x_{n+1}} \right) = \Omega_{n+1}\iv{x_{i}}\v{x_{n+1}}
\end{eqnarray}
where $\Omega_{n}\conj $ and $\Omega_{n+1} $ are non-zero transition coefficients. The first term vanishes by 
the inductive hypothesis. 
Since $H$ is Hermitian, we have 
\begin{eqnarray}\label{eq:AppA:Hconj} 
 \iv{x_{i}}H\v{x_{n}} = \left( \iv{x_{n}}H\v{x_{i}} \right)\conj = \left[ \iv{x_{n}} \left( \Omega_{i}\conj \v{x_{i-1}}+ \Omega_{i+1} \v{x_{i+1}} \right)\right]\conj = 0
 \end{eqnarray}
By inductive hypothesis, both averages vanish. Note that this relies on the hypothesis that $ i \neq n-1 $, which 
guarantees that $ \v{x_{n}} $ and $ \v{x_{i+1}} $ are not the same. Setting this equation equal to 
(\ref{eq:AppA:Hdot}) gives 
\begin{eqnarray}\label{eq:AppA:QED}
 \Omega_{n+1}\iv{x_{i}}\v{x_{n+1}} = 0
\end{eqnarray}
Since $ \Omega_{n+1} $ is a non-zero coefficient, $ \iv{x_{i}}\v{x_{n+1}}$ must be 0. This completes the proof by induction.

\section{Derivation of $N_{3}$}\label{app:N3Der}
Here we normalize the state $\v{y_{3}}$ by computing its magnitude. We recall
\begin{eqnarray}\label{eq:AppD:eq4} 
 \v{y_{3}} \!=\! \left(b_{11}c_{10}\!+\!c_{11}b_{10}\right)\v{100}
 \!+\!\left(a_{11}c_{01}\!+\!c_{11}a_{10}\right)\v{010}
 \!+\!\left(a_{11}b_{01}\!+\!b_{11}a_{01}\right)\v{001}
\end{eqnarray}
The normalizing factor $N_{3}$ is equal to the magnitude of the right side, such that
\begin{eqnarray}\label{eq:AppD:eq5} \nonumber
 N_{3}^{2}= 2b^{2}c^{2}+2Re\left(b_{11}c_{11}^{*}b_{10}^{*}c_{10}\right) \\ \nonumber 
 +2a^{2}c^{2} + 2Re\left(a_{11}^{*}c_{11}a_{10}c_{01}^{*}\right) \\ 
 +2a^{2}b^{2} + 2Re\left(a_{11}b_{11}^{*}b_{01}^{*}a_{01}\right)
\end{eqnarray}
We can rewrite this in terms of magnitudes and phases
\begin{eqnarray}
 N_{3}^{2}= 2\left[a^{2}b^{2}\left(\cos\phi_{1}+1\right)+a^{2}c^{2}\left(\cos\phi_{2}+1\right)+b^{2}c^{2}\left(\cos\phi_{3}+1\right)\right]
\end{eqnarray}
Finally, we obtain
\begin{eqnarray}\label{eq:AppD:N3} 
N_{3} = 2\sqrt{a^{2}b^{2}\cos^{2}\frac{\phi_{1}}{2}+a^{2}c^{2}\cos^{2}\frac{\phi_{2}}{2}+a^{2}+b{2}c^{2}\cos^{2}\frac{\phi_{3}}{2}}
\end{eqnarray}
\section{Derivation of $\v{x_{4}}$}\label{app:X4Der}
The fourth basis state is derived by applying Eq. (\ref{eq:AA:YBas}) to $\v{y_{3}}$

\begin{eqnarray}\label{eq:AppE:eq1} 
  \v{y_{4}} = H'\v{y_{3}}-\left(\frac{N_{3}}{N_{2}} \right)^{2}\v{y_{2}}
\end{eqnarray}
The first term is found as
\begin{eqnarray}\label{eq:AppE:eq2} \nonumber
 H'\v{y_{3}}= \left[a_{11}\left( b^{2}+c^{2}\right)+b_{11}a_{01}b_{01}^{*}+c_{11}a_{10}c_{01}^{*}\right]\v{011} \\ \nonumber
 +\left[ a_{11}a_{01}^{*}b_{01}+b_{11}\left(a^{2}+c^{2}\right)+ c_{11}b_{10}c_{10}^{*}\right]\v{101} \\ \nonumber 
 +\left[a_{11}a_{10}^{*}c_{01}+ b_{11}b_{10}^{*}c_{10}+ c_{11}\left(a^{2}+b^{2}\right)\right] \v{110} \\ \nonumber
 + [a_{00}\left(b_{11}c_{10}+c_{11}b_{10}\right)+b_{00}\left(a_{11}c_{01}+c_{11}a_{10}\right) \\ 
 +c_{00}\left(a_{11}b_{01}+b_{11}a_{01}\right)]\v{000}
\end{eqnarray}
Since $\v{y_{2}}= a_{11}\v{011}+b_{11}\v{101}+c_{11}\v{110}$, only the $a_{11}$, $b_{11}$, and $c_{11}$ 
terms change when the projection is subtracted off. As a result, we obtain
\begin{eqnarray}\label{eq:AppE:eq4} \nonumber
 \v{y_{4}} = \left[a_{11}\left( b^{2}+c^{2}-\frac{N_{3}^{2}}{N_{2}^{2}}\right)+b_{11}a_{01}b_{01}^{*}+c_{11}a_{10}c_{01}^{*}\right]\v{011} \\ \nonumber
 +\left[ a_{11}a_{01}^{*}b_{01}+b_{11}\left(a^{2}+c^{2}-\frac{N_{3}^{2}}{N_{2}^{2}}\right)+ c_{11}b_{10}c_{10}^{*}\right]\v{101} \\ 
 +\left[a_{11}a_{10}^{*}c_{01}+ b_{11}b_{10}^{*}c_{10}+ c_{11}\left(a^{2}+b^{2}-\frac{N_{3}^{2}}{N_{2}^{2}}\right)\right] \v{110}
\end{eqnarray}
Finally, we normalize and express it in terms of $\phi$ phases. We obtain
\begin{eqnarray}\label{eq:AppE:eq5} \nonumber
 N_{4}^{2} &=& a^{2}\left(b^{4} +c^{4}+\frac{N_{3}^{4}}{N_{2}^{4}}+2b^{2}c^{2} -2b^{2}\frac{N_{3}^{2}}{N_{2}^{2}}-2c^{2}\frac{N_{3}^{2}}{N_{2}^{2}  }\right) \\ \nonumber 
 &+&a^{2}b^{4}+a^{2}c^{4}+2a^{2}b^{2}\cos\phi_{1}\left( b^{2}+c^{2}-\frac{N_{3}^{2}}{N_{2}^{2}}\right) \\ \nonumber
 &+&2a^{2}c^{2}\cos\phi_{2}\left( b^{2}+c^{2}-\frac{N_{3}^{2}}{N_{2}^{2}}\right) + 2a^{2}b^{2}c^{2}\cos\left( \phi_{1}+\phi_{2}\right) \\ \nonumber
 &+&a^{4}b^{2}+b^{2}\left(a^{4} +c^{4}+\frac{N_{3}^{4}}{N_{2}^{4}}+2a^{2}c^{2} -2a^{2}\frac{N_{3}^{2}}{N_{2}^{2}}-2c^{2}\frac{N_{3}^{2}}{N_{2}^{2}  }\right) \\ \nonumber
 &+&b^{2}c^{4}+2a^{2}b^{2}\cos\phi_{1}\left(a^{2}+c^{2}-\frac{N_{3}^{2}}{N_{2}^{2}}\right) \\ \nonumber 
 &+&2a^{2}b^{2}c^{2}\cos\left(\phi_{1}+\phi_{3} \right)+2b^{2}c^{2}\cos\phi_{3}\left(a^{2}+c^{2}-\frac{N_{3}^{2}}{N_{2}^{2}}\right) \\ \nonumber 
 &+&a^{4}c^{2}+b^{4}c^{2}+c^{2}\left(a^{4} +b^{4}+\frac{N_{3}^{4}}{N_{2}^{4}}+2a^{2}b^{2} -2a^{2}\frac{N_{3}^{2}}{N_{2}^{2}}-2b^{2}\frac{N_{3}^{2}}{N_{2}^{2}  }\right) \\ \nonumber
 &+&2a^{2}b^{2}c^{2}\cos\left(\phi_{2}+\phi_{3} \right)+2a^{2}c^{2}\cos\phi_{2}\left( a^{2}+b^{2}-\frac{N_{3}^{2}}{N_{2}^{2}}\right) \\
 &+&2b^{2}c^{2}\cos\phi_{3}\left(a^{2}+b^{2}-\frac{N_{3}^{2}}{N_{2}^{2}}\right)
\end{eqnarray}
From first cube splitting condition given by Eq. (\ref{eq:CG:Cond1'}) we have
\begin{eqnarray}\label{eq:AppE:eq6} 
\cos\left( \phi_{2}+\phi_{3}\right) &=& -\cos\phi_{1} \\
\cos\left( \phi_{1}+\phi_{3}\right) &=& -\cos\phi_{2}\\
\cos\left( \phi_{1}+\phi_{2}\right) &=& -\cos\phi_{3}
\end{eqnarray}
As a result, all terms of the form 
$2abc\cos\left( \phi_{i}+\phi_{j}\right)$ cancel out with a term of the form $2abc\cos\phi_{k}$, 
simplifying the normalization factor to
\begin{eqnarray}\label{eq:AppE:eq7} \nonumber
N_{4}^{2} &=& 2\left( a^{2}b^{4}+a^{4}b^{2}+a^{2}c^{4}+a^{4}c^{2}+b^{2}c^{4}+b^{4}c^{2}+3a^{2}b^{2}c^{2}\right)   \\ \nonumber
&+&2\left(a^{2}b^{4}+a^{4}b^{2} \right)\cos\phi_{1} + 2\left(a^{2}c^{4}+a^{4}c^{2} \right)\cos\phi_{2}+2\left(b^{2}c^{4}+b^{4}c^{2} \right)\cos\phi_{3} \\ \nonumber
&+& 2a^{2}b^{2}c^{2}\left(\cos\phi_{1}+\cos\phi_{2}+\cos\phi_{3} \right) \\ \nonumber
&-&4\left[a^{2}b^{2}\left(1+\cos\phi_{1} \right)+a^{2}c^{2}\left(1+\cos\phi_{2} \right)+b^{2}c^{2}\left(1+\cos\phi_{3} \right)\right]\frac{N_{3}^{2}}{N_{2}^{2}} \\
&+& \left( a^{2}+b^{2}+c^{2}\right)\frac{N_{3}^{4}}{N_{2}^{4}}
\end{eqnarray}
This equation can be further simplified by expressing everything in terms of $N_{2}$ and $N_{3}$.
We obtain
\begin{eqnarray}\label{eq:AppE:eq8} \nonumber
N_{4}^{2}=N_{2}^{2}N_{3}^{2}-2N_{3}^{2}\frac{N_{3}^{2}}{N_{2}^{2}}+N_{2}^{2}\frac{N_{3}^{4}}{N_{2}^{4}} = \frac{N_{3}^{2}}{N_{2}^{2}}\left(N_{2}^{4}-N_{3}^{2} \right)
\end{eqnarray}

\end{document}